\documentclass{fac}
\usepackage{graphicx}
\usepackage{amssymb}
\usepackage{amsmath}
\ifprodtf \else \usepackage{latexsym}\fi

\newcommand\black{\ensuremath{\blacktriangleright}}
\newcommand\white{\ensuremath{\vartriangleright}}

\newif\ifamsfontsloaded
\ifprodtf
  \newcommand\whbl{\white\kern-.1em--\kern-.1em\black}
  \newcommand\blwh{\black\kern-.1em--\kern-.1em\white}
  \newcommand\blbl{\black\kern-.1em--\kern-.1em\black}
  \newcommand\whwh{\white\kern-.1em--\kern-.1em\white}
  \amsfontsloadedtrue
\else
  \checkfont{msam10}
  \iffontfound
    \IfFileExists{amssymb.sty}
      {\usepackage{amssymb}\amsfontsloadedtrue
       \newcommand\whbl{\white\kern-.125em--\kern-.125em\black}%
       \newcommand\blwh{\black\kern-.125em--\kern-.125em\white}%
       \newcommand\blbl{\black\kern-.125em--\kern-.125em\black}%
       \newcommand\whwh{\white\kern-.125em--\kern-.125em\white}}
      {}
  \fi
\fi

\newcommand{\getClientId}[1]{getClientId(#1)}
\newcommand{\getWSOOntology}[1]{getWSOOntology(#1)}
\newcommand{\getInputParameters}[1]{getInputParameters(#1)}
\newcommand{\getQoS}[1]{getQoS(#1)}
\newcommand{\getWSOReq}[2]{getWSOReq(#1,#2)}

\newcommand{\getWSOReqFromWSOI}[1]{getWSOReq(#1)}
\newcommand{\getState}[1]{getState(#1)}
\newcommand{\getAAs}[1]{getAAs(#1)}
\newcommand{\getOutputParameters}[1]{getOutputParameters(#1)}
\newcommand{\createNewAAName}[1]{createNewAAName(#1)}
\newcommand{\getWSOI}[2]{getWSOI(#1, #2)}

\newcommand{\getAA}[2]{getAA(#1, #2)}
\newcommand{\getAAFromC}[3]{getAA(#1, #2, #3)}
\newcommand{\getAAQoS}[1]{getAAQoS(#1)}
\newcommand{\getInputParametersFromAA}[1]{getInputParameters(#1)}
\newcommand{\getOutputParametersFromAA}[1]{getOutputParameters(#1)}
\newcommand{\getAAState}[1]{getAAState(#1)}
\newcommand{\getAAName}[1]{getAAName(#1)}
\newcommand{\getWSOIId}[1]{getWSOIId(#1)}
\newcommand{\getWSFromAA}[1]{getWS(#1)}

\newcommand{\getWSesFromWSOI}[1]{getWSes(#1)}
\newcommand{\getWSFromWSOI}[2]{getWS(#1, #2)}
\newcommand{\getWSFromC}[3]{getWS(#1, #2, #3)}
\title[A Formal Model of QoS-Aware Web Service Orchestration Engine]
      {A Formal Model of QoS-Aware Web Service Orchestration Engine}

\author[Yong Wang]
    {Yong Wang\\
     College of Computer Science and Technology,\\
     Beijing University of Technology, Beijing, China\\
     }

\correspond{Yong Wang, Pingleyuan 100, Chaoyang District, Beijing, China.
            e-mail: wangy@bjut.edu.cn}

\pubyear{2011}
\pagerange{\pageref{firstpage}--\pageref{lastpage}}

\begin{document}
\label{firstpage}

\makecorrespond

\maketitle

\begin{abstract}
QoS-aware applications can satisfy not only the functional requirements of the customers, but also the QoS requirements. QoS-aware Web Service orchestration translates the QoS requirements of the customers into those of its component Web Services. In a system viewpoint, we discuss issues on QoS-aware Web Service orchestration and design a typical QoS-aware Web Service orchestration engine called QoS-WSOE. More importantly, we establish a formal model of QoS-WSOE based on actor systems theory. Within the formal model, we use a three-layered pyramidal structure to capture the requirements of the customers with a concept named QoS-Aware WSO Service, characteristics of QoS-WSOE with a concept named QoS-Aware WSO System, and structures and behaviors of QoS-WSOE with a concept named QoS-Aware WSO Behavior. Conclusions showing that a system with QoS-Aware WSO Behavior is a QoS-Aware WSO System and further can provide QoS-Aware WSO Service are drawn.
\end{abstract}

\begin{keywords}
Web Services; Web Service Orchestration; Web Service Orchestration Engine; Actor Systems; QoS; Formal Model
\end{keywords}

\section{Introduction}

Web Service (WS) is a new distributed component which emerged about ten years ago, which uses WSDL\cite{WSDL} as its interface description language, SOAP\cite{SOAP} as its communication protocol and UDDI\cite{UDDI} as its directory service. Because WS uses the Web as its provision platform, it is suitable to be used to develop cross-organizational business integrations.

Cross-organizational business processes are usual forms in e-commerce that orchestrate some business activities into a workflow. WS Orchestration (WSO) provides a solution for such business process based on WS technologies, hereby representing a business process where business activities are modeled as component WSes.

From a WS viewpoint, WSO provides a workflow-like pattern to orchestrate existing WSes to create a new composite WS, and embodies the added values of WS. In particular, We use the term WSO, rather than another term -- WS Composition, because there are also other WS composition patterns, such as WS Choreography (WSC) \cite{WS-CDL}. However, about WSC and the relationship of WSO and WSC\cite{WSO and WSC 1}, we do not explain more, because it is not the focus of this paper.

In this paper, we focus on WSO, exactly, the QoS-aware WSO engine (runtime of WSO) and its formal model. A QoS-aware WSO enables the customers to be satisfied with not only their functional requirements, but also their QoS requirements, such as performance requirements, reliability requirements, security requirements, etc. A single execution of a WSO is called a WSO instance. A QoS-aware WSO engine provides runtime supports for WSOs with assurance of QoS implementations. These runtime supports include lifetime operation on a WSO instance, queue processing for request from the customers and incoming message delivery to a WSO instance.

WS and WSO are with a continuously changing and evolving environment. The customers, the requirements of the customers, and the component WSes are all changing dynamically. To assure safe adaptation to dynamically changing and evolving requirements, it is important to have a rigorous semantic model of the system: the component WSes, the WSO engine that provides WSO instance management and invocation of the component WSes, the customer accesses, and the interactions among these elements. Using such a model, designs can be analyzed to clarify assumptions that must be met for correct operation.

We design a typical QoS-aware WSO engine, called QoS-WSOE in this paper. An architecture of QoS-WSOE is given, and more importantly, a formal model of QoS-WSOE is established based on actor systems theory\cite{Actor 1}\cite{Actors}\cite{Actor reasoning}. In the formal model, we introduce the notion of \textbf{QoS-Aware WSO Service}, \textbf{QoS-Aware WSO System}, and \textbf{QoS-Aware WSO Behavior}. And we draw conclusions that: (1) if a QoS-aware WSO engine is a \textbf{QoS-Aware WSO System}, then it provides \textbf{QoS-Aware WSO Service} for customers; (2) if a QoS-aware engine has \textbf{QoS-Aware WSO Behavior}, then it is a \textbf{QoS-Aware WSO System} and further provides \textbf{QoS-Aware WSO Service}.

This paper is organized as follows. In section 2, we introduce the related works. The actor computational model and the three layered pyramidal architecture are introduced in section 3. We illustrate a WS composition example called BuyingBooks in section 4. We design QoS-WSOE in section 5. And in section 6, formal model of QoS-WSOE is established. Finally, we conclude our works and point out future works.

\section{Related Works}

The main efforts on WSO of the industry are trying to establish a uniform WSO description language specification, such as the early WSFL\cite{WSFL}, XLANG\cite{XLANG}, and lately converged WS-BPEL\cite{WS-BPEL}. Such WSO description languages based on different mathematical models have constructs to model invocation of WSes, manipulate information transferred between WSes, control execution flows of these activities and inner transaction processing mechanisms. The WSO description language can be used to define various WSOs under different requirements and acts as a so-called meta language. WSOs described by such meta languages actually are pure texts and must be enabled by the meta language interpreter called WSO engine, such as the famous open source ActiveBPEL\cite{Active BPEL}, ReSpecT tuple centres based WS-BPEL engine\cite{BPELEngine1}, a multi-agent system based WS-BPEL engine\cite{BPELEngine2} and event-driven architecture based WS-BPEL engine\cite{BPELEgine3}.

In industry, there are many research works to give the meta languages correctness verifications based on different theoretical tools\cite{Survey}. \cite{Form BPEL PNet} formalizes WS-BPEL\cite{WS-BPEL} with Petri-Net. \cite{Form BPEL PA} uses process algebra to give WS-BPEL a theoretical foundation. \cite{Form WSO Calculus} establishes a calculus to verify correctness of WS-BPEL. Semantics and verifications of WS-BPEL are researched in \cite{Form BPEL 2}. \cite{WSOF} uses a kind of formal specification to orchestrate WSes. Unlike these formalizations, our formal model focuses on the correctness of the WSO engine, that is, correctness of a runtime system of WSO, but not correctness of the meta languages.

QoS-aware WSO engine can process not only the functional requirements modeled by the WSO description language, but also the QoS requirements of the customers. For example, a WSO must complete within three hours and the cost running some WSO must be under twenty dollars. A QoS-aware WSO translates the QoS requirements of the customers into QoS requirements of component WSes. Such a translation is accomplished by the implementation of the so-called QoS-aware Service Selection Algorithm (QSSA). There are many kind of such service selection algorithms, such as \cite{Service Selection Algo 1}, which uses the so-called local service selection approach and the global allocation approach based on integer programming, and another one in \cite{Service Selection Algo 2}, which models the service selection problem (SSP) in two ways: one defines SSP as a multi-dimensional multi-choice 0-1 knapsack problem (MMKP) based on combinatorial model and the other captures SSP as a multi-constraint optimal path problem (MCOP) based on graph model, and gives heuristic algorithms of the two ways. We do not research SSP itself, but use the implementation of a QSSA above as a component of our WSO engine QoS-WSOE.

Note that QoS-aware WSO implies that the WSes involved in this WSO also must be QoS-aware. In this paper, we assume that the WSes are all QoS-aware. About QoS of WS, the readers please refer to \cite{Web Service QoS 1} and \cite{Web Service QoS 2}. \cite{Web Service QoS 2} also involves implementations of QoS-aware WSes.

Actor\cite{Actor 1}\cite{Actors} is a basic concurrent computing model and can be used in reasoning about open distributed systems\cite{Actor reasoning}. In \cite{Actor Resource} and \cite{Actor QoS Middleware}, actors are used to reason about middleware of resource management in distributed computing and even QoS-aware middleware. Our works follow the works above, especially \cite{Actor QoS Middleware}. That is, in this paper, we adopt the way for formalization of open distributed systems in \cite{Actor QoS Middleware} which uses a pyramidal refinement to capture the concepts of customer requirements, system requirements, system behaviors and their relationships. Our works focus on QoS-aware WSO engine, a software system different to the multimedia resource management middleware in \cite{Actor QoS Middleware} with respect to different QoS aspects and QoS management, different system functions and architecture, different system components and behaviors. That is, the formal model in this paper is a different one from that in \cite{Actor QoS Middleware}.

\section{Actors and the Three Layered Pyramidal Architecture}

An actor\cite{Actor 1}\cite{Actors} is a basic concurrent computation unit which encapsulates a set of local states, a control thread and a set of local computations. It has a unique mail address and maintains a mail box to receive messages sent by other actors. Through processing the messages stored in the main box sequentially, an actor computes locally and blocks when its mail box is empty.

During processing a message from its mail box, an actor may perform three candidate actions: (1)(\textbf{send} )sending messages asynchronously to other actors; (2)(\textbf{create}) creating new actors with new behaviors; (3)(\textbf{ready}) becoming ready again to process the next message from the mail box or block if the mail box is empty.


Note that synchronization\cite{Actor Syn} can also be achieved among actors. Also there are many works to abstract at a high-level from aspects of distributed computing, such as policy management\cite{Actor Cust and Compo}, interaction policies\cite{Actor Interaction}, resource management\cite{Actor Resource}, communication and coordination of agents\cite{Actor Agent}, worldwide computing\cite{Actor WWW}, etc.


The actor computational model can be used to reason about the behavior of a computer system, especially the behavior of a distributed system\cite{Actor Resource}. \cite{Actor QoS Middleware} uses actors to reason about the behavior of QoS-aware distributed middleware and presents a so-called pyramidal structure to capture the concepts of customer requirements, system requirements, system behavior and their relationships. Our works are greatly inspired by this pyramidal structure and are with the similar goal but different system. That is, our QoS-aware WSO Engine called QoS-WSOE has similar requirements, but different behavior, with the QoS-aware distributed middleware in \cite{Actor QoS Middleware}, just because they are different systems in nature.

\section{A Bookstore WSO in the BuyingBooks Example}

In this section, we give a so-called BuyingBooks example for the scenario of cross-organizational business process integration and use a so-called BookStore WSO to illustrate some related concepts, such as WSO, activity, etc. And we use the BookStore WSO to explain the formal model we established in the following.

\subsection{A BuyingBooks Example}

\begin{figure*}
  \centering
  \includegraphics{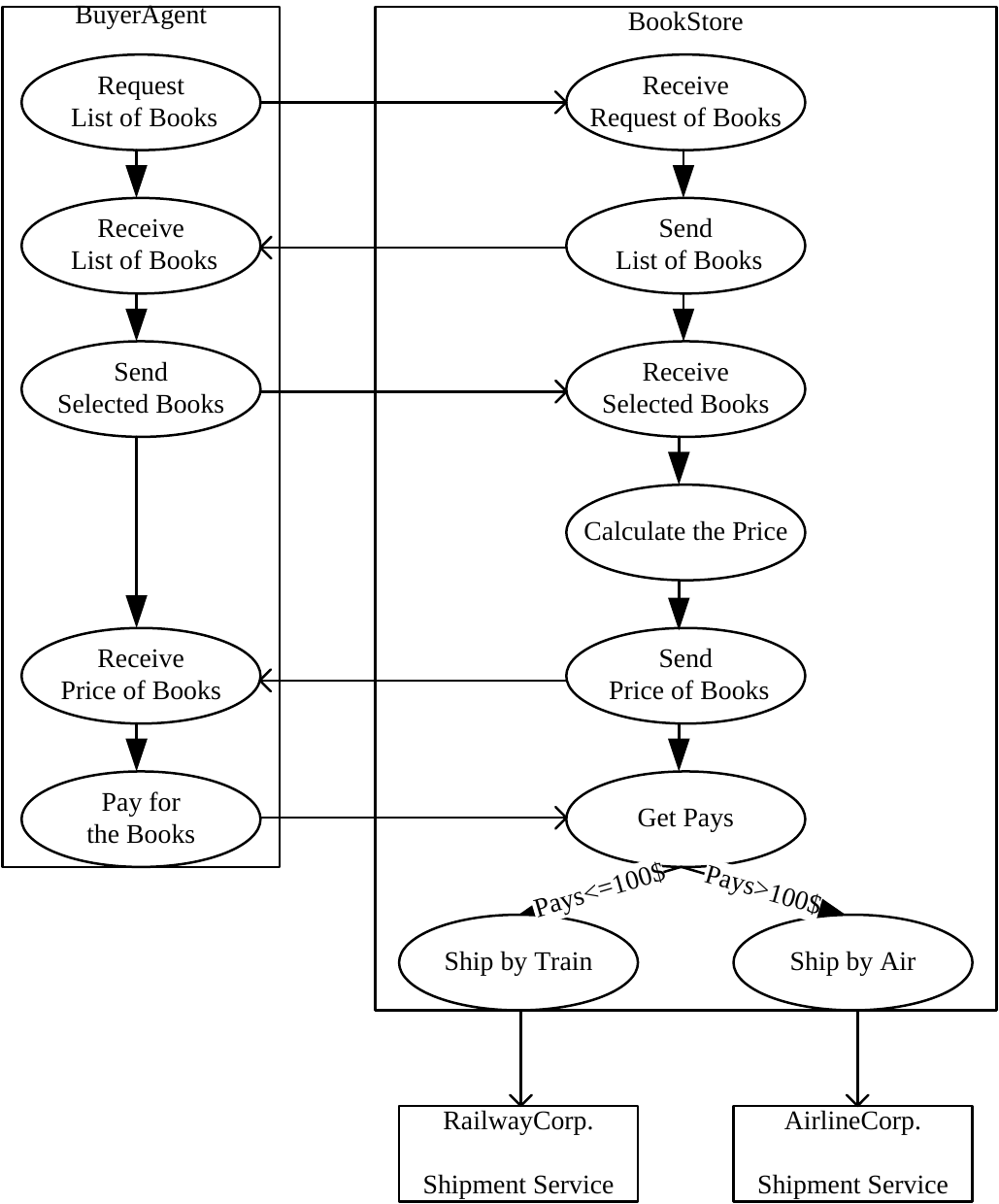}
  \caption{The BuyingBooks Example.}
  \label{Fig.BuyingBooksExample}
\end{figure*}

A further example is BuyingBooks as Fig.\ref{Fig.BuyingBooksExample} shows. We use this BuyingBooks example throughout this paper to illustrate concepts and mechanisms in WS Composition.

In Fig.\ref{Fig.BuyingBooksExample}, there are four organizations: BuyerAgent, BookStore, RailwayCorp, and AirlineCorp. And each organization has one business process. Exactly, there are two business processes, the business processes in RailwayCorp and AirlineCorp are simplified as just WSes for simpleness without loss of generality. We introduce the business process of BookStore as follows, and the process of BuyerAgent can be understood as contrasts.

\begin{enumerate}
  \item The BookStore receives request of list of books from the buyer through BuyerAgent.
  \item It sends the list of books to the buyer via BuyerAgent.
  \item It receives the selected book list by the buyer via BuyerAgent.
  \item It calculates the price of the selected books.
  \item It sends the price of the selected books to the buyer via BuyerAgent.
  \item It gets pays for the books from the buyer via BuyerAgent.
  \item If the pays are greater than 100\$, then the BookStore calls the shipment service of AirlineCorp for the shipment of books.
  \item Otherwise, the BookStore calls the shipment service of RailwayCorp for the shipment of book. Then the process is ended.
\end{enumerate}

Each business process is implemented by a WSO, for example, the BookStore WSO and BuyerAgent WSO implement BookStore process and BuyerAgent process respectively. Each WSO invokes external WSes through its activities directly. And each WSO is published as a WS to receive the incoming messages.

\subsection{The Bookstore WSO}

The BookStore WSO described by WS-BPEL is given as follows.

--------------------------------------------------------------

$\langle$process name="BookStore"

\quad targetNamespace="http://example.wscs.com

\quad\quad /2011/ws-bp/bookstore"...$\rangle$

\quad $\langle$partnerLinks$\rangle$

\quad\quad $\langle$partnerLink name="BSAndBA"... /$\rangle$

\quad\quad $\langle$partnerLink name="BSAndRC"... /$\rangle$

\quad\quad $\langle$partnerLink name="BSAndAC"... /$\rangle$

\quad $\langle$/partnerLinks$\rangle$

\quad $\langle$variables$\rangle$

\quad\quad $\langle$variable name="RequestListofBooks"

\quad\quad\quad messageType="lns:requestListofBooks"/$\rangle$

\quad\quad $\langle$variable name="RequestListofBooksResponse"

\quad\quad\quad messageType="lns:requestListofBooksResponse"/$\rangle$

\quad\quad $\langle$variable name="ListofBooks"

\quad\quad\quad messageType="lns:listofBooks"/$\rangle$

\quad\quad $\langle$variable name="ListofBooksResponse"

\quad\quad\quad messageType="lns:listofBooksResponse"/$\rangle$

\quad\quad $\langle$variable name="SelectListofBooks"

\quad\quad\quad messageType="lns:selectListofBooks"/$\rangle$

\quad\quad $\langle$variable name="SelectListofBooksResponse"

\quad\quad\quad messageType="lns:selectListofBooksResponse"/$\rangle$

\quad\quad $\langle$variable name="Price"

\quad\quad\quad messageType="lns:price"/$\rangle$

\quad\quad $\langle$variable name="PriceResponse"

\quad\quad\quad messageType="lns:priceResponse"/$\rangle$

\quad\quad $\langle$variable name="Pays"

\quad\quad\quad messageType="lns:pays"/$\rangle$

\quad\quad $\langle$variable name="PaysResponse"

\quad\quad\quad messageType="lns:paysResponse"/$\rangle$

\quad\quad $\langle$variable name="ShipmentByTrain"

\quad\quad\quad messageType="lns:shipmentByTrain"/$\rangle$

\quad\quad $\langle$variable name="ShipmentByTrainResponse"

\quad\quad\quad messageType="lns:shipmentByTrainResponse"/$\rangle$

\quad\quad $\langle$variable name="ShipmentByAir"

\quad\quad\quad messageType="lns:shipmentByAir"/$\rangle$

\quad\quad $\langle$variable name="ShipmentByAirResponse"

\quad\quad\quad messageType="lns:shipmentByAirResponse"/$\rangle$

\quad $\langle$/variables$\rangle$

\quad $\langle$sequence$\rangle$

\quad\quad $\langle$receive

\quad\quad\quad partnerLink="BSAndBA"

\quad\quad\quad portType="lns:bookStore4BuyerAgent-

\quad\quad\quad\quad Interface"

\quad\quad\quad operation="opRequestListofBooks"

\quad\quad\quad variable="RequestListofBooks"

\quad\quad\quad createInstance="yes"$\rangle$

\quad\quad $\langle$/receive$\rangle$

\quad\quad $\langle$invoke

\quad\quad\quad partnerLink="BSAndBA"

\quad\quad\quad portType="bns:buyAgent4BookStore-

\quad\quad\quad\quad Interface"

\quad\quad\quad operation="opReceiveListofBooks"

\quad\quad\quad inputVariable="ListofBooks"

\quad\quad\quad outputVariable="ListofBooksResponse"$\rangle$

\quad\quad $\langle$/invoke$\rangle$

\quad\quad $\langle$receive

\quad\quad\quad partnerLink="BSAndBA"

\quad\quad\quad portType="lns:bookStore4BuyerAgent-

\quad\quad\quad\quad Interface"

\quad\quad\quad operation="opSelectListofBooks"

\quad\quad\quad variable="SelectListofBooks"$\rangle$

\quad\quad $\langle$/receive$\rangle$

\quad\quad $\langle$reply

\quad\quad\quad partnerLink="BSAndBA"

\quad\quad\quad portType="lns:bookStore4BuyerAgent-

\quad\quad\quad\quad Interface"

\quad\quad\quad operation="opSelectListofBooks"

\quad\quad\quad variable="SelectListofBooksResponse"$\rangle$

\quad\quad $\langle$/reply$\rangle$

\quad\quad $\langle$!--inner activity: calculate the price

\quad\quad\quad of selected books--$\rangle$

\quad\quad $\langle$invoke

\quad\quad\quad partnerLink="BSAndBA"

\quad\quad\quad portType="bns:buyAgent4BookStore-

\quad\quad\quad\quad Interface"

\quad\quad\quad operation="opReceivePrice"

\quad\quad\quad inputVariable="Price"

\quad\quad\quad outputVariable="PriceResponse"$\rangle$

\quad\quad $\langle$receive

\quad\quad\quad partnerLink="BSAndBA"

\quad\quad\quad portType="lns:bookStore4BuyerAgent-

\quad\quad\quad\quad Interface"

\quad\quad\quad operation="opPays" variable="Pays"$\rangle$

\quad\quad $\langle$/receive$\rangle$

\quad\quad $\langle$reply

\quad\quad\quad partnerLink="BSAndBA"

\quad\quad\quad portType="lns:bookStore4BuyerAgent-

\quad\quad\quad\quad Interface"

\quad\quad\quad operation="opPays"

\quad\quad\quad variable="PaysResponse"$\rangle$

\quad\quad $\langle$if$\rangle$$\langle$condition$\rangle$ getVariable('Price')

\quad\quad\quad\quad $\langle$ 100 $\langle$/condition$\rangle$

\quad\quad\quad $\langle$invoke

\quad\quad\quad\quad partnerLink="BSAndAC"

\quad\quad\quad\quad portType="ans:airlineCorp4BookStore-

\quad\quad\quad\quad\quad Interface"

\quad\quad\quad\quad operation="opShipmentByAir"

\quad\quad\quad\quad inputVariable="ShipmentByAir"

\quad\quad\quad\quad outputVariable="ShipmentByAirResponse"$\rangle$

\quad\quad\quad $\langle$else$\rangle$

\quad\quad\quad\quad $\langle$invoke

\quad\quad\quad\quad\quad partnerLink="BSAndRC"

\quad\quad\quad\quad\quad portType="rns:railwayCorp4BookStore-

\quad\quad\quad\quad \quad\quad Interface"

\quad\quad\quad\quad\quad operation="opShipmentByTrain"

\quad\quad\quad\quad\quad inputVariable="ShipmentByTrain"

\quad\quad\quad\quad\quad outputVariable="ShipmentByTrain-

\quad\quad\quad\quad\quad\quad Response"$\rangle$

\quad\quad\quad $\langle$/else$\rangle$$\langle$/if$\rangle$

\quad $\langle$/sequence$\rangle$

$\langle$/process$\rangle$

--------------------------------------------------------------

There are several receive-reply activity pairs and several invoke activities in the BookStore WSO. The QoS requirements are not included in the WS-BPEL description, because these need an extension of WS-BPEL and are out of the scope of this paper. In the request message from the BuyerAgent WSO, the QoS requirements, such as the whole execution time threshold and the additional charges, can also be attached, not only the functional parameters.

Another related specification is the WSDL description of the interface WS for BuyingBooks WSO. Because we focus on WS composition, this WSDL specification is omitted.

\section{Architecture of A Typical QoS-Aware WSO Engine, QoS-WSOE}

In this section, we firstly analyze the requirements of a WSO Engine. And then we discuss problems about QoS management of WS and define the QoS aspects used in this paper. Finally, we give the architecture of QoS-WSOE and discuss the state transition of a WSO instance.

\subsection{Requirements for A WSO Engine and QoS Management of WS}

As the introduction above says, a WSO description language, such as WS-BPEL, has:

 \begin{itemize}
   \item basic constructs called atomic activities to model invocation to an external WS, receiving invocation from an external WS and reply to that WS, and other inner basic functions;
   \item information and variables exchanged between WSes;
   \item control flows called structural activities to orchestrate activities;
   \item other inner transaction processing mechanisms, such as exception definitions and throwing mechanisms, event definitions and response mechanisms.
 \end{itemize}

Therefore, a WSO described by WS-BPEL is a program with WSes as its basic function units and must be enabled by a WSO engine. An execution of a WSO is called an instance of that WSO. The WSO engine can create a new WSO instance according to information included in a request of a customer via the interface WS (Note that a WSO is encapsulated as a WS also.) of the WSO. Once a WSO is created, it has a thread of control to execute independently according to its definition described by a kind of description language, such as WS-BPEL. During its execution, it may create activities to interact with WSes outside and also may do inner processings, such as local variable assignments. When it ends execution, it replies to the customer with its execution outcomes.

In order to provide the adaptability of a WSO, the bindings between its activities and WSes outside are not direct and static. That is, WSes are classified according to ontologies of specific domains and the WSes belonging to the same ontology have same functions and interfaces, and different access points and different QoS. To make this possible, from a system viewpoint, a name and directory service -- UDDI\cite{UDDI} is necessary. All WSes with access information and QoS information are registered into a UDDI which classifies WSes by their ontologies to be discovered and invoked in future. UDDI should provide multi interfaces to search WSes registered in for its users, for example, a user can get information of specific set of WSes by providing a service ontology and specific QoS requirements via an interface of the UDDI.

The above mechanisms make QoS-aware service selection possible. In a QoS-aware WSO engine, after a new WSO instance is created, the new WSO instance firstly selects its component WSes according to the QoS requirements provided by the customer and ontologies of component WSes defined in the description file of the WSO by WS-BPEL.

About QoS of a WS\cite{Web Service QoS 1}\cite{Web Service QoS 2}, there are various QoS aspects, such as performance QoS, security QoS, reliability QoS, availability QoS, and so on. In this paper, we use a cost-effective QoS approach. That is, cost QoS is used to measure the costs of one invocation of a WS while response time QoS is used to capture effectiveness of one invocation of a WS. In the following, we assume all WSes are aware of cost-effective QoS.

\subsection{Architecture of QoS-WSOE}

\begin{figure*}
  \centering
  \includegraphics{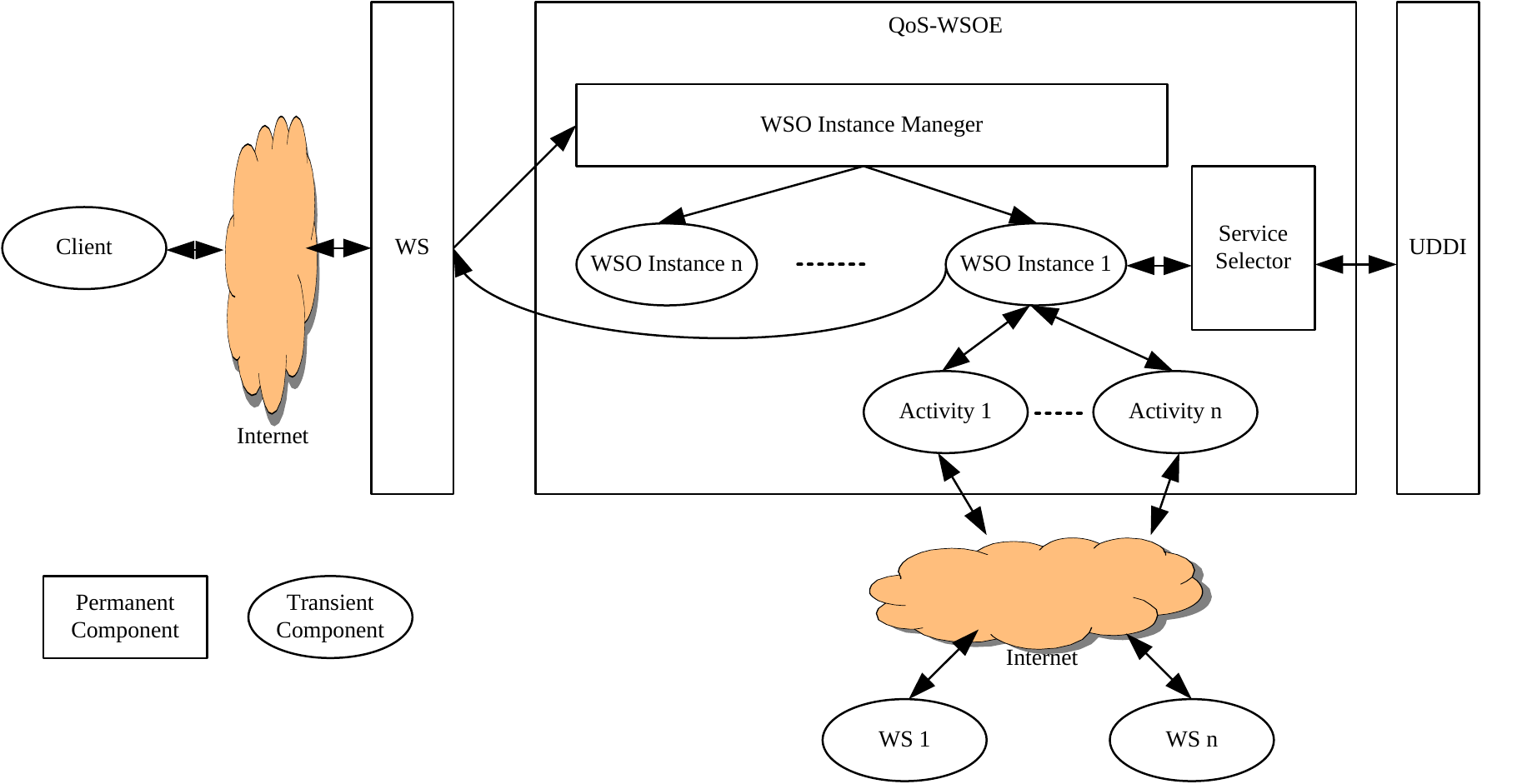}
  \caption{Architecture of QoS-WSOE.}
  \label{Fig.1}
\end{figure*}

According to the requirements of a WSO engine discussed above, the architecture of QoS-WSOE is given as Fig.\ref{Fig.1} shows.

In the architecture of QoS-WSOE, there are external components, such as Client, WS of a WSO, UDDI and component WSes, and inner components, including WSO Instance Manager, WSO Instances, Activities, and Service Selector. Among them, WS of a WSO, UDDI, WSO Instance Manager and Service Selector are permanent components and Client, component WSes, WSO Instances, Activities are transient components. Component WSes are transient components since they are determined after a service selection process is executed by Service Selector.

Through a typical requirement process, we illustrate the functions and relationships of these components.

\begin{enumerate}
  \item A Client submits its requests including the WSO ontology, input parameters and QoS requirements to the WS of a WSO through SOAP protocol.
  \item The WS transmits the requirements from a SOAP message sent by the Client to the WSO Instance Manager using private communication mechanisms.
  \item The WSO Instance Manager creates a new WSO Instance including its Activities and transmits the input parameters and the QoS requirements to the new instance.
  \item The instance transmits ontologies of its component WSes and the QoS requirements to the Service Selector to perform a service selection process via interactions with a UDDI. If the QoS requirements can not be satisfied, the instance replies to the Client to deny this time service.
  \item If the QoS requirements can be satisfied, each activity in the WSO Instance is bound to an external WS.
  \item The WSO Instance transmits input parameters to each activity for an invocation to its binding WS.
  \item After the WSO Instance ends its execution, that is, every invocation to its component WSes by activities in the WSO Instance is returned, the WSO Instance returns the execution outcomes to the Client.
\end{enumerate}

\subsection{WSO Instance -- An Execution of A WSO}

An execution of a WSO is called a WSO instance (WSOI). A WSOI is created when the WSO Instance Manager receive a new request (including the functional parameters and the QoS requirements).

\begin{figure}
  \centering
  \includegraphics{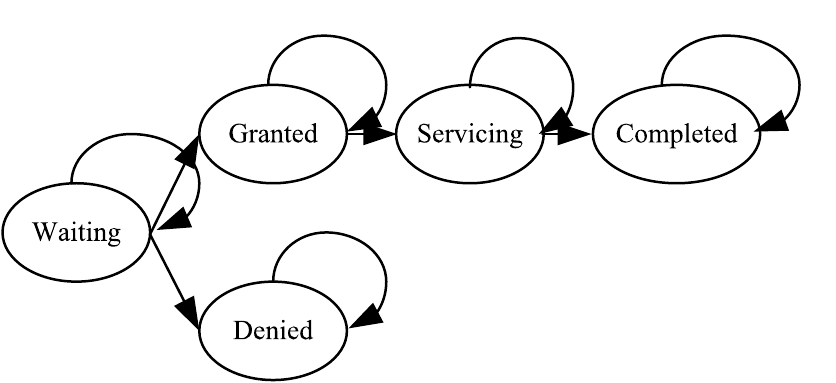}
  \caption{State Transitions of A WSO Instance.}
  \label{Fig.WSOI}
\end{figure}

As Fig.\ref{Fig.WSOI} shows, once a WSOI is created, the WSOI is in the \emph{Waiting} state. Then the WSO Instance Manager requires the Service Selector to perform a service selection process to select suitable component services. If the QoS requirements can not be satisfied, the WSOI transmits to \emph{Denied} state, otherwise, the WSOI transmits to the \emph{Granted} state. When the WSOI starts to execute, that is, the first activity is executed, the WSOI is in the state of \emph{Servicing} and executes according to the definition of the WSO. Finally, the WSOI ends to execute, that is, every activity of the WSOI is completed, the WSOI is with the state of \emph{Completed}.

\begin{figure}
  \centering
  \includegraphics{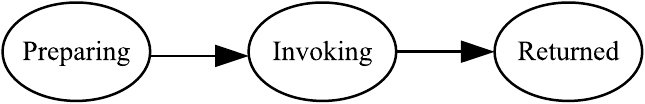}
  \caption{State Transitions of An Activity.}
  \label{Fig.AA}
\end{figure}

Every instance of an activity (we do not distinguish the uses of an activity and an activity instance) is included in a WSOI. When a WSOI is granted, all activities are in the state of \emph{Preparing}. In addition, an activity is with the state of \emph{Invoking} when it is the executing turn of this activity. Once this activity is completed, that is, it gets the outcomes of the invocation to the external WS, it is in the \emph{Returned} state. The state transitions is illustrated in Fig.\ref{Fig.AA}.

\begin{figure*}
  \centering
  \includegraphics{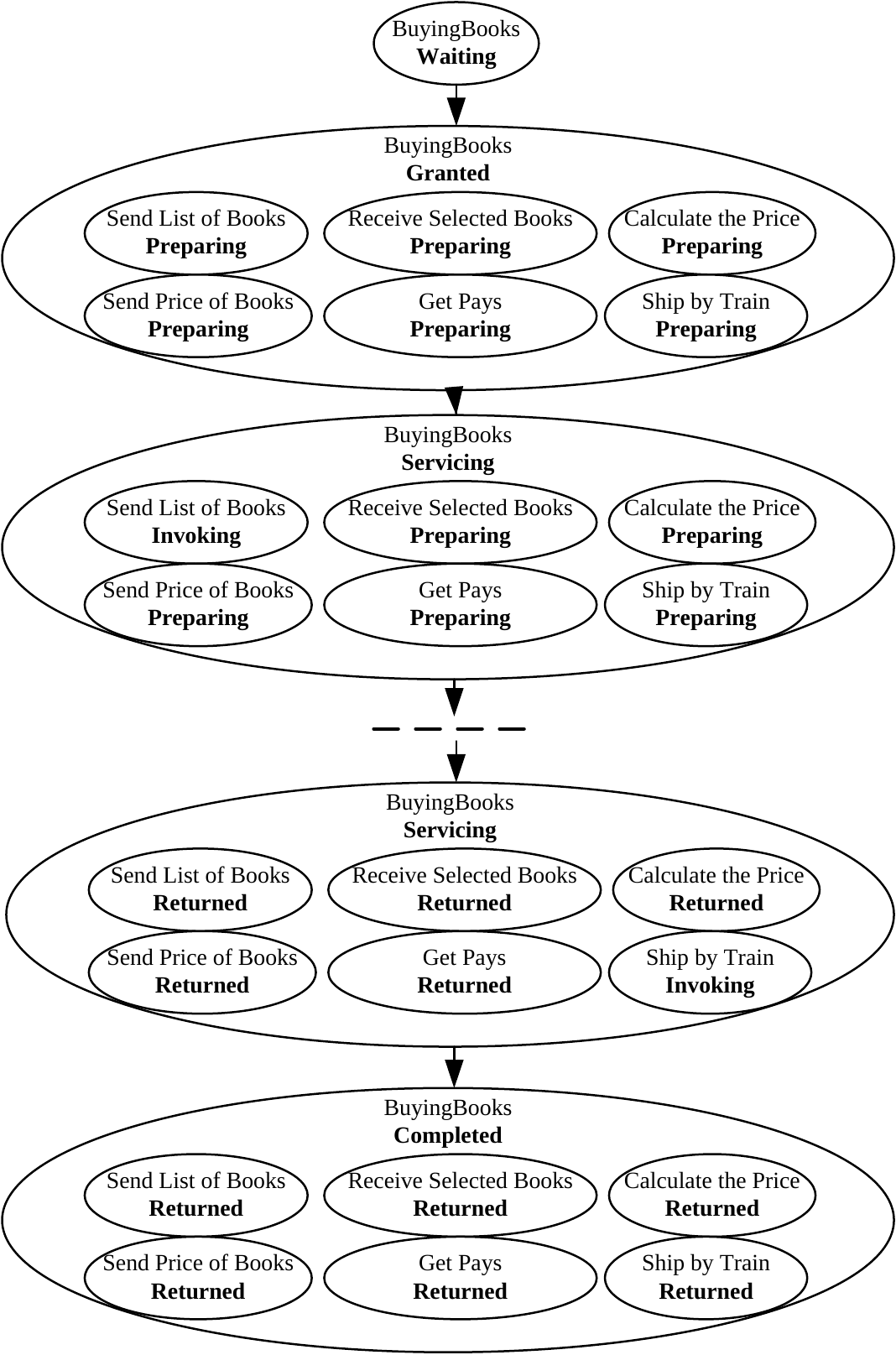}
  \caption{State Transitions of A BookStore WSOI and Its Activities.}
  \label{Fig.BookStore}
\end{figure*}

Based on the state transitions of a WSOI and an activity as shown in Fig.\ref{Fig.WSOI} and Fig.\ref{Fig.AA}, we can get the state transitions of a BookStore WSOI as Fig.\ref{Fig.BookStore} illustrates when the QoS requirements of the request are satisfied.

\subsection{The Glossary and the Symbols Used in This Paper}

Components in Fig.\ref{Fig.1} are all implemented as actors. Some of these actors used in the following are Client Actor (CA), WSO Instance Manager Actor (WSOIM), WSO Instance Actor (WSOI), Activity Actor (AA), Service Selector Actor (SS), component WS Actor (WS).

We follow the symbol convention in \cite{Actor Resource} and \cite{Actor QoS Middleware}. Some are following, and others are introduced when we need.

$C$ denotes a system configuration.

$Cast(C)$ denotes the set of names of actors existing in $C$.

$getState(C, a)$ denotes the state of actor $a$ in $C$.

$getA(C, a, t)$ gets the value of tag $t$ of actor $a$ in $C$.

$setA(C, a, t, v)$ sets the value of tag $t$ of actor $a$ in $C$ with a value $v$.

$\langle a:s\rangle$ denotes an actor with a name $a$ and a state $s$.

$b\lhd MessageType(paras)@a$ denotes a message with a type $MessageType$ sent from actor $a$ to actor $b$ with optional parameters $paras$.

$\langle a:s\rangle [, a\lhd M]\xrightarrow[effect]{trigger}\langle a:s'\rangle, MC \quad if \quad condition$ denotes a transition rule, where $\langle a:s\rangle$ is an actor with name $a$ and state $s$, $a\lhd M$ is a message to actor $a$ with content $M$, $s'$ is the new state of actor $a$, $trigger$ is the event triggering this rule, $effect$ is the effect of this rule such as $setA(C, a, t, v)$, $MC$ is the set of (possible empty) messages sent in this rule, $condition$ is the condition of occurrence of this rule.

$\tau:C\xrightarrow{l}C'$ denotes a transition, $C=source(\tau)$ is the source configuration, $C'=target(\tau)$ is the target configuration, and $l$ is the label of the transition rule applied.

$\pi=[C_i\xrightarrow{l_i}C_{i+1}|i\in \textbf{Nat}]$ denotes a computation path, which is a possibly infinite sequence of transitions.

$\tau_i=C_i\xrightarrow{l_i}C_{i+1}$ is called $i$th transition of $\pi$ or $i$th stage of $\pi$.

\section{Formal Model of QoS-WSOE}

In this section, we will introduce the formal model of QoS-WSOE. Firstly, we introduce actors and the symbols used in this section. Then we establish a pyramidal formal structure of QoS-WSOE, including \textbf{QoS-Aware WSO Service} to model the requirements of the customers, \textbf{QoS-Aware WSO System} to capture the properties of QoS-aware WSO engine, and \textbf{QoS-Aware WSO Behavior} to capture the behaviors of QoS-aware WSO engine.

\subsection{QoS-Aware WSO Service}

Let us use a WSO ontology $WSO$ range over the set of WSO ontologies $WSOOntologies$, and let a request from the customer $WSOReq$ range over the set of requests $WSOReqs$.

\textbf{Definition 4.1. (WSO Request)} A WSO Request actor $WSOReq$ is defined as a 4-tuples $WSOReq=\langle ClientId=\alpha_{cl},$ $ WSOOntology=WSO,$ $ InputParameters=ips,$ $ QoS=qos\rangle$, where $\alpha_{cl}$ is the client ID, $WSO$ is the WSO ontology requested, $ips$ is the input parameters of the WSO ontology, and $qos$ is the QoS requirements.

\textbf{Definition 4.2. (WSO Request Functions)}To manipulate a WSO Request $WSOReq$, we define the following functions:

(1)$\getClientId{WSOReq}=\alpha_{cl}$ denotes the function to get the Client ID;

(2)$\getWSOOntology{WSOReq}=WSO$ denotes the function to get the WSO ontology;

(3)$\getInputParameters{WSOReq}=ips$ denotes the function to get the input parameters;

(4)$\getQoS{WSOReq}=qos$ denotes the function to get the QoS requirements;

(5)$\getWSOReq{C}{\alpha_{cl}}=WSOReq$ denotes the function to get a WSO Request with a Client ID $\alpha_{cl}$ from a configuration $C$.

\textbf{Definition 4.3. (QoS-Aware WSO Service)}A system $S$ provides a QoS-Aware WSO Service over $WSOOntologies$ and $WSOReqs$ iff for every configuration $C$ of $S$, if there is an undelivered request $WSOReq$ in $C$, then along any path $\pi$ of $C$, exactly one of the following properties holds:

(1)there is a unique transition $\tau$ in $\pi$ where $WSOReq$ is accepted for service and the QoS requirements $qos$ of $WSOReq$ can be satisfied;

(2)or there is a unique transition $\tau$ in $\pi$ where $WSOReq$ is rejected, only because the QoS requirements $qos$ of $WSOReq$ can not be satisfied when $WSOReq$ arrives.

\subsection{QoS-Aware WSO System}

When a $WSOReq$ message is sent to the WSOIM, the WSOIM creates a new WSOI. Now we give the definition of WSOI.

\textbf{Definition 4.4. (WSOI)}A WSOI is defined as a 4-tuples $WSOI=\langle WSORequest=WSOReq,\\ WSOIState=state,\\ AAs=AAs,\\ OutputParameters=ops\rangle$, where $WSOReq$ is the WSO Request, $state$ is the state of the WSOI, $ops$ denotes the output parameters of the WSOI, and $AAs$ is activity actors included in the WSOI. $state$ ranges over the set $\{Waiting, Granted, Denied, Servicing, Completed\}$, where $Waiting$ denotes that a WSOI is created and is waiting for further processing, $Granted$ denotes that a WSOI is accepted and the QoS requirements can be satisfied, $Denied$ denotes that a WSO is rejected because the QoS requirements can not be satisfied, $Servicing$ denotes that a WSOI is under running, and $Completed$ denotes that a WSOI is completed in running and the QoS requirements are satisfied.

\textbf{Definition 4.5. (WSOI Functions)}To manipulate a WSOI, the following functions are defined:

(1)$\getWSOReqFromWSOI{WSOI}=WSOReq$ denotes the function to get the WSO Request contained in the WSO instance $WSOI$;

(2)$\getState{WSOI}=state$ denotes the function to get the state of the WSO instance $WSOI$;

(3)$\getAAs{WSOI}=AAs$ denotes the function to get the set of activity actors contained in the WSO instance $WSOI$;

(4)$\getOutputParameters{WSOI}=ops$ denotes the function to get the output parameters of the WSO instance $WSOI$;

(5)$\createNewAAName{WSOI}$ creates a new AA name;

(6)$\getWSOI{C}{\alpha_{cl}}=WSOI$ denotes the function to get a WSO instance $WSOI$ with a Client ID $\alpha_{cl}$ from the configuration $C$, since a WSO instance and a WSO Request are with 1:1 relation.

We define AA as follows.

\textbf{Definition 4.6. (AA)}An AA is defined as a 7-tuples $AA=\langle AAName=aaName,\\ WSOIId=\alpha_{cl},\\ QoS=qos_{AA},\\ InputParameters=ips_{AA},\\ OutParameters=ops_{AA},\\ AAState=state_{AA},\\ WS=ws\rangle$, where $aaName$ is the name of the AA, $\alpha_{cl}$ denotes the ID of a WSO instance, $qos_{AA}$ is the QoS requirements of the AA, $ips_{AA}$ is the input parameters of the AA, $ops_{AA}$ is the output parameters of the AA, $state_{AA}$ is the state of the AA, and $ws$ is the WS bound after a service selection process. $state_{AA}$ ranges over $\{Preparing, Invoking, Returned\}$, where $Preparing$ denotes that an AA is created by a WSOI and is waiting for invoking a WS, $Invoking$ denotes that an AA is now invoking a WS, and $Returned$ denotes that the invocation from an AA to a WS is completed.

\textbf{Definition 4.7. (AA Functions)}To manipulate an AA, we define the following functions:

(1)$\getAA{WSOI}{aaName}=AA$ denotes the function to get an activity actor $AA$ with a name $aaName$ contained in the WSO instance $WSOI$;

(2)$\getAAFromC{C}{\alpha_{cl}}{aaName}=AA$ denotes the function to get an activity actor $AA$ with a name $aaName$ contained in the WSO instance $\alpha_{cl}$ from a configuration $C$;

(3)$\getAAQoS{AA}=qos_{AA}$ denotes the function to get the QoS requirements of an AA $AA$;

(4)$\getInputParametersFromAA{AA}=ips_{AA}$ denotes the function to get the input parameters of an AA $AA$;

(5)$\getOutputParametersFromAA{AA}=ops_{AA}$ denotes the function to get the output parameters of an AA $AA$;

(6)$\getAAState{AA}=state_{AA}$ denotes the function to get the state of an AA $AA$;

(7)$\getAAName{AA}=aaName$ denotes the function to get the name of an AA $AA$;

(8)$\getWSOIId{AA}=\alpha_{cl}$ denotes the function to get the WSOI ID of an AA $AA$;

(9)$\getWSFromAA{AA}=ws$ denotes a function to get the WS bound to an AA $AA$.

WS is defined as follows.

\textbf{Definition 4.8. (WS)}A WS is determined by a pair $\langle WSOId=\alpha_{cl}, WSId=aaName\rangle$, where $\alpha_{cl}$ denotes the ID the a WSOI, and $aaName$ denotes the name $aaName$ of AA that the WS bound. Since a WS is out of the management domain of the WSOI, we assume that a WS is always able to process the invocation of its customers, such as an AA.

\textbf{Definition 4.9. (WS Functions)}To manipulate A WS, we define the following functions:

(1)$\getWSFromAA{AA}=ws$ denotes the function to get the WS bound of an AA $AA$;

(2)$\getWSesFromWSOI{WSOI}=wses$ denotes the function to get WSes included in a WSOI $WSOI$;

(3)$\getWSFromWSOI{WSOI}{aaName}=ws$ denotes the function to get the WS bound to the AA with a name $aaName$ contained in the WSO instance $WSOI$.

(4)$\getWSFromC{C}{\alpha_{cl}}{aaName}=ws$ denotes the function to get a WS $ws$ bound to an AA with a name $aaName$ contained in the WSO instance $\alpha_{cl}$ from a configuration $C$;

\textbf{Definition 4.10. (QoS Requirements of WS and WSOI)}We use a cost-effectiveness couple $ResponseTime\times Cost$ to capture the QoS requirements of WS and WSOI. If a WSOI has $n$ WSes: $ws_1,...,ws_n$, then there is the following relation between the WSOI and its WSes:

$Cost_{WSOI}=\sum\limits_{i=1}^n (Cost_{ws_i})$,

$ResponseTime_{WSOI}=\max\limits_{i=1}^n (ResponseTime_{ws_i})$.

Since all AAs included in a WSOI are executed in parallel by default, we can get the above equations if we ignore the causalities of messages sending among the AAs.

\textbf{Definition 4.11. (QoS-Allocate Function)}The QoS-allocate function translates the QoS requirements of a WSOI into the QoS requirements of its WSes implemented by some kind of QSSA, such as in \cite{Web Service QoS 1} and \cite{Web Service QoS 2}, with an effect of binding AAs in the WSOI to WSes selected. The QoS-allocate function is defined as follows:

$ResponseTime_{WSOI}\times Cost_{WSOI}\rightarrow$

$\bigsqcup\limits_{i=1}^n(ResponseTime_{WS_i}\times Cost_{WS_i})$ where $WSOI$ has $n$ WSes and $1\leq i\leq n$.

\textbf{Definition 4.12. (AA-InputParametersGenerate Function)}The AA-InputParametersGenerate function maps the input parameters of a WSOI to those of AAs. The AA-InputParametersGenerate is defined as follows:

$ips_{WSOI}\rightarrow \biguplus\limits_{i=1}^n(ips_{WS_i})$ where $WSOI$ has $n$ WSes and $1\leq i\leq n$.

\textbf{Definition 4.13. (WSOI-OutputParametersGenerate Function)}The WSOI-OutputParametersGenerate function maps the output parameters of AAs to those of WSOI. The WSOI-OutputParametersGenerate is defined as follows:

$\biguplus\limits_{i=1}^n(ops_{WS_i})\rightarrow ops_{WSOI}$ where $WSOI$ has $n$ WSes and $1\leq i\leq n$.

\textbf{Definition 4.14. (WSOI Constraints $\phi_{WSOI}$)}The WSOI Constraints $\phi_{WSOI}$ includes:

(1)A system $S$ satisfies $\phi_{WSOI}$,$\phi_{WSOI}(S)$, just if for every computation $\pi$ of $S$, $\phi_{WSOI}(\pi)$ holds.

(2)$\phi_{WSOI}(\pi)$holds, just if for every transition $\tau$ of $\pi$, $\phi_{WSOI}(\tau)$ holds, and

(2-a)A WSOI in some configuration $C$ with the state $Waiting$ will eventually be granted or be denied.

$
(\forall i\in \textbf{Nat})(C=source(\pi(i)) \\ \wedge \getState{\getWSOI{C}{\alpha_{cl}}}=Waiting)\nonumber\\
\Rightarrow (\exists j\in \textbf{Nat})
(C'=source(\pi(i+j+1)) \\ \wedge \getState{\getWSOI{C'}{\alpha_{cl}}}\in\{Granted, Denied\})
$

(2-b)A WSOI in some configuration $C$ with the state $Granted$ will eventually be in servicing and be completed.

$
(\forall i\in \textbf{Nat})(C=source(\pi(i))\wedge \getState{\getWSOI{C}{\alpha_{cl}}}=Granted)\nonumber\\
\Rightarrow (\exists j,j'\in \textbf{Nat})(C'=source(\pi(i+j+1)),\\ C''=source(\pi(i+j'+1)) \\ \wedge \getState{\getWSOI{C'}{\alpha_{cl}}}=Servicing \\ \wedge \getState{\getWSOI{C''}{\alpha_{cl}}}=Completed)\nonumber
$

(3)$\phi_{WSOI}(\tau)$ holds for $\tau:C\rightarrow C'$, just if

(3-a)For a WSOReq in some configuration $C$, once the corresponding WSOI is created, then the WSOReq and all AAs in the WSOI are constant.

$
(\forall WSOReq\in WSOReqs\cap Cast(C)) \\ (\getWSOReqFromWSOI{\getWSOI{C}{WSOReq}}\nonumber\\
=\getWSOReqFromWSOI{\getWSOI{C'}{WSOReq}}\\ \wedge(\forall AA\in \getAAs{\getWSOI{C}{WSOReq}}\\
(\forall AA'\in \getAAs{\getWSOI{C'}{WSOReq}}\\ \getWSOIId{AA}=\getWSOIId{AA'})\\ \wedge
(\exists AA''\in \getAAs{\getWSOI{C'}{WSOReq}}\\ \getAAName{AA}=\getAAName{AA''})))
$

(3-b)For any WSOreq in some configuration $C$, if the state of the corresponding WSOI is $Waiting$, then the WSOI may be in a state $Waiting$, or $Granted$, or $Denied$; if the state of the corresponding WSOI is $Denied$, then the WSOI will keep in the state $Denied$; if the corresponding WSOI is in a state $Granted$, the it will still be in a state $Granted$ or in a new state $Servicing$; if the corresponding WSOI is still in the state $Servicing$, the it will still be in a state $Servicing$ or in a new state $Completed$; if the state of the corresponding WSOI is $Completed$, then the WSOI will keep in the state $Completed$.

$(\forall WSOReq\in WSOReqs\cap Cast(C))$

$\getState{\getWSOI{C}{WSOReq}}=Waiting \\ \Rightarrow  \getState{\getWSOI{C'}{WSOReq}} \\ \in\{Waiting, Granted, Denied\}$,

and

$\getState{\getWSOI{C}{WSOReq}}=Denied \\ \Rightarrow  \getState{\getWSOI{C'}{WSOReq}}=Denied$,

and

$\getState{\getWSOI{C}{WSOReq}}=Granted \\ \Rightarrow  \getState{\getWSOI{C'}{WSOReq}} \\ \in\{Granted, Servicing\}$,

and

$\getState{\getWSOI{C}{WSOReq}}=Servicing \\ \Rightarrow  \getState{\getWSOI{C'}{WSOReq}} \\ \in\{Servicing, Completed\}$,

and

$\getState{\getWSOI{C}{WSOReq}}=Completed \\ \Rightarrow  \getState{\getWSOI{C'}{WSOReq}}=Completed$.

(3-c)If the corresponding WSOI of any WSOReq in some configuration $C$ is denied, then the WSes included in the WSOI will always be $nil$, that is, after a service selection process executed for this WSOI, the QoS requirements of the WSOReq can not be satisfied.

$
\getState{\getWSOI{C}{WSOReq}}=Denied\\
\Rightarrow (\forall ws\in \getWSesFromWSOI{\getWSOI{C}{WSOReq}}\\
 \wedge \getWSesFromWSOI{\getWSOI{C'}{WSOReq}})\\
  ws=nil
$

\textbf{Definition 4.15. (WSOI-WS Constraints $\phi_{WW}$)}A system $S$ satisfies $\phi_{WW}$,$\phi_{WW}(S)$, just if for every computation $\pi$ of $S$, $\phi_{WW}(\pi)$ holds, iff for any WS included in the corresponding WSO of any WSOReq in some configuration $C$ is not $nil$, then the WSOI will in a state $Granted$, or $Servicing$, or $Completed$, that is,

$
 (\forall ws\in \getWSesFromWSOI{\getWSOI{C}{WSOReq}}) ws\neq nil\nonumber\\
 \Rightarrow  \getState{\getWSOI{C}{WSOReq}}\\ \in \{Granted, Servicing, Completed\}\nonumber
$

\textbf{Definition 4.16. (QoS-Aware WSO System)}A system $S$ is a QoS-Aware WSO System, with respect to request $WSOReq$ in $WSOReqs$, the above functions including QoS-Allocate function, AA-InputParametersGenerate function, WSOI-OutputParametersGenerate function, WSO Request functions, WSOI functions, AA functions and WS functions, iff (1)$S$ satisfies $\phi_{WSOI}(S)$ and $\phi_{WW}(S)$; (2)for $C\in S$, if there is an undelivered request $WSOReq$ with parameters $\langle \alpha_{cl}, WSO, ips, qos\rangle$, then along any path $\pi$ of $C$ there is a unique stage $i$ transition: $\pi(i)=C\rightarrow C'$ to process $WSOReq$, and there is a newly created WSO instance $WSOI=getWSOI(C, \alpha_{cl})$, with parameters:

\begin{itemize}
  \item $\getWSOReqFromWSOI{WSOI}=WSOReq$;
  \item $\getState{WSOI}=Waiting$;
  \item $\getOutputParameters{WSOI}=nil$;
  \item for $\forall AA\in \getAAs{WSOI}$,
  \begin{itemize}
    \item $\getAAQoS{AA}=nil$;
    \item $\getInputParametersFromAA{AA}=nil$;
    \item $\getOutputParametersFromAA{AA}=nil$;
    \item $\getAAState{AA}=Preparing$;
    \item $\getWSOIId{AA}=\alpha_{cl}$;
    \item $\getWSFromAA{AA}=nil$;
    \item $\getAAName{AA}\\ =\createNewAAName{WSOI}$.
  \end{itemize}
\end{itemize}

\textbf{Theorem 4.1. (Service2System)}If a system $S$ is a QoS-Aware WSO System as Definition 4.16 shows, then this system $S$ provides QoS-Aware WSO Service as defined in Definition 4.3.

\begin{proof}
We firstly see that if there is an undelivered request $WSOReq$ for any $C$ in $S$, according to definition 4.16, a new WSO instance $WSOI$ is created and $C\xrightarrow{+} C_{Waiting}$. Then $C_{Waiting}\xrightarrow{+} C_{Granted}$ or $C_{Waiting}\xrightarrow{+} C_{Denied}$ according to Definition 4.14, if in $C_{Denied}$, a reply is sent to the customer to reject the request; if in $C_{Granted}$, then $C_{Granted}\xrightarrow{+} C_{Servicing}$ and $C_{Servicing}\xrightarrow{+} C_{Completed}$. Thus, $WSOReq$ is assured to be just accepted or just rejected.

QoS-Allocate function, $\phi_{WSOI}$, $\phi_{WW}$ and Definition 4.8 guarantee that the QoS requirements are satisfied in case that $WSOReq$ is accepted.
\end{proof}

\subsection{QoS-Aware WSO Behavior}

In this section, we will specify the behavior of actors in Fig.\ref{Fig.1}, including their states, messages exchanged among them and their transition rules. Then we define the concept of QoS-Aware WSO Behavior and draw two conclusions.

WSOI, AA and WS are defined in the above section including their states. Now we will explain WSOIM and SS. The functions of WSOIM are creating WSOIs when incoming requests from CA arrive. We assume that WSOIM are always ready for servicing. And the functions of SS are providing QoS-aware service selection processing for WSOIs and SS is also always ready for servicing.

We define the messages exchanged among these actors as Table.\ref{Table.1} illustrates.

\begin{center}
\begin{table*}
  \caption{Request-Reply and Notification Messages Definition}
  \begin{tabular}{@{}ll@{}}
   \hline
   Request Messages\hspace{3cm}                                & Reply Messages \\
   \hline
   $AA\lhd invoke()@WSOI$                                      & $WSOI\lhd invokeAck()@AA$ \\
   \hline
   $WS\lhd invoke(ips)@AA$                                     & $AA\lhd invokeReply(ops)@WS$ \\
   \hline
   $SS\lhd select(qos, WSO)@WSOI$                              & $WSOI\lhd selectReply(WSOI=[State=Denied])@SS$ \\
                                                               & $WSOI\lhd selectReply(WSOI=[State=Granted])@SS$ \\
   \hline
   $WSOIM\lhd wsoReq(\alpha_{cl}, WSO, qos, ips)@CA$           &$CA\lhd granted(WSO, qos)@WSOI$ \\
                                                               & $CA\lhd completed(WSO, qos, ops)@WSOI$ \\
                                                               & $CA\lhd denied(WSO, qos)@WSOI$ \\
   \hline
   Notifications                                               &\\
   \hline
   $WSOI\lhd notify(AA=[State=Returned])@AA$                   &\\
   \hline
  \end{tabular}
  \label{Table.1}
\end{table*}
\end{center}

The followings are the transition rules for WSOIM and WSOI.

(1)Rule for WSOIM creating new WSOI: when the WSOIM receives a incoming request message, then a new WSOI is created and a service selection processing message will be sent from the new WSOI to the SS.

$
\langle WSOIM:WSOIMState\rangle, \\ WSOIM\lhd wsoReq(\alpha_{cl}, WSO, qos, ips)@CA\nonumber\\
\xrightarrow[effect_{newWSOI}]{} \langle WSOIM:WSOIMState'\rangle,\\ SS\lhd select(qos, WSO)@WSOI\nonumber
$.

Where $effect_{newWSOI}\\ =new(WSOI);setA(wsoiU)$, and

$
wsoiU=WSOI\{ \\ \getWSOReqFromWSOI{WSOI}=WSOReq,\\
\getState{WSOI}=Waiting,\\
\getOutputParameters{WSOI}=nil,\\
\forall AA\in \getAAs{WSOI}\{\\
\getAAQoS{AA}=nil,\\
 \getInputParametersFromAA{AA}=nil, \\
 \getOutputParametersFromAA{AA}=nil,\\
 \getAAState{AA}=Preparing, \\
 \getWSOIId{AA}=\alpha_{cl},
\getWSFromAA{AA}=nil,\\
 \getAAName{AA}=\createNewAAName{WSOI}\} \}
$

(2)Rules for WSOI processing reply from SS: after a service selection process is executed by the SS, a reply message from the SS to the WSOI is sent. If the QoS requirements can not be satisfied, then the WSOI is in a new state $Denied$ and a \emph{denied} reply message is sent from the WSOI to the customer. If the QoS requirements can be satisfied, then the WSOI is in a new state $Granted$ and a \emph{granted} reply message is sent from the WSOI to the customer and a \emph{invoke} message will be sent to any AA included in the WSOI from the WSOI.

$
\langle WSOI:Waiting\rangle, \\ WSOI\lhd selectReply(WSOI=[State=Denied])@SS\\
\rightarrow \langle WSOI:Denied\rangle,\\ CA\lhd denied(WSO, qos)@WSOI
$.

$
\langle WSOI:Waiting\rangle, WSOI \\ \lhd selectReply(WSOI=[State=Granted])@SS\\
\xrightarrow[effect_{grantedWSOI}]{} \langle WSOI:Granted\rangle, \\ CA\lhd granted(WSO, qos)@WSOI,\\
(\forall AA\in \getAAs{WSOI}) AA\lhd invoke()@WSOI
$.

Where $effect_{grantedWSOI}=setA(grantedWSOIU)$, and

$
grantedWSOIU=WSOI\{\\
\getWSOReqFromWSOI{WSOI}=WSOReq, \\
\getState{WSOI}=Granted,\\
\getOutputParameters{WSOI}=nil, \\
\forall AA\in \getAAs{WSOI}\{\\
\getAAQoS{AA}=QoSAllocate(AA), \\
\getInputParametersFromAA{AA}\\
=AA-InputParametersGenerate(AA), \\
\getOutputParametersFromAA{AA}=nil, \\
\getAAState{AA}=Preparing, \\
\getWSOIId{AA}=\alpha_{cl},\\
\getWSFromAA{AA}=ServiceSelection(AA), \\
\getAAName{AA}=aaName\} \}
$,

where $QoSAllocate(AA)$ denotes the requirements after a service selection processing by SS, $AA-InputParametersGenerate(AA)$ denotes input parameters generation of AA from the WSOI, and $ServiceSelection(AA)$ denotes the WS bound to the AA after a service selection processing by SS.

(3)Rule for WSOI processing reply from AA: when a WSOI in a state $Granted$ or $Servicing$ receives a reply message \emph{invokeAck} from an AA, then it will be in a new state $Servicing$.

$
\langle WSOI:\{Granted, Servicing\}\rangle, \\
WSOI\lhd invokeAck()@AA \\
\rightarrow \langle WSOI:Servicing\rangle
$.

(4)Rules for WSOI processing notification from AA: when a WSOI in a state $Servicing$ receives a notification message from an AA, if all AAs included in the WSOI are in state $Returned$, then the WSOI will be in a new state $Completed$ and a \emph{completed} reply message will be sent from the WSOI to the customer.

$
\langle WSOI:Servicing\rangle, \\
WSOI\lhd notify(AA=[State=Returned])@AA\nonumber\\
\xrightarrow[effect_{completedWSOI}]{} \langle WSOI:Completed\rangle, \\
CA\lhd completed(WSO, qos, ops)@WSOI
$.

Where $(\forall AA\in \getAAs{WSOI}) \\ \getAAState{AA}=Returned$ and $effect_{completedWSOI}=setA(completedWSOIU)$, and

$
completedWSOIU=WSOI\{\\
\getWSOReqFromWSOI{WSOI}=WSOReq, \\
\getState{WSOI}=Completed,\\
\getOutputParameters{WSOI}\\
=WSOI-OutputParametersGenerate(WSOI),\\
\forall AA\in \getAAs{WSOI}\{\\
\getAAQoS{AA}=qos_{AA}, \\
\getInputParametersFromAA{AA}=ips_{AA},\\
\getOutputParametersFromAA{AA}\\
=ResultsFromWS(AA), \\
\getAAState{AA}=Returned, \\
\getWSOIId{AA}=\alpha_{cl},\\
getWS(AA)=ws, \\
getAAName(AA)=aaName\} \}
$,

where $ResultsFromWS(AA)$ denotes the results after a WS invocation, $WSOI-OutputParametersGenerate(WSOI)$ denotes output parameters generation of WSOI from the AAs.

When a WSOI in a state $Servicing$ receives a notification message from an AA, if there is an AA included in the WSOI is not in state $Returned$, then the WSOI will keep the state $Servicing$.

$
\langle WSOI:Servicing\rangle, \\ WSOI\lhd notify(AA=[State=Returned])@AA\nonumber\\
\rightarrow \langle WSOI:Servicing\rangle
$.

Where $(\exists AA\in \getAAs{WSOI}) \getAAState{AA}\neq Returned$.

The following is the transition rule for SS to process request from WSOI: when the SS receives a \emph{select} request message from a WSOI, then a reply message \emph{selectReply} will be sent to the WSOI for denying the request or granting the request.

$
\langle SS:SSState\rangle, \\SS\lhd select(qos, WSO)@WSOI \nonumber\\
\rightarrow \langle SS:SSState'\rangle, \\(WSOI\lhd selectReply\\\quad(WSOI=[State=Denied])@SS\\ \quad or
WSOI\lhd selectReply\\\quad(WSOI=[State=Granted])@SS)
$.

The followings are the transition rules for AA.

(1)Rule for AA to process request from WSOI: an AA in a state $Preparing$ receives a request message \emph{invoke} from its WSOI, then it will be in a new state $Invoking$, send an \emph{invokeAck} reply message to the WSOI, and send an \emph{invoke} request message to its binding WS.

$
\langle AA:Preparing\rangle, \\ AA\lhd invoke()@WSOI \nonumber\\
\rightarrow \langle AA:Invoking\rangle, \\ WSOI\lhd invokeAck()@AA,  WS\lhd invoke(ips)@AA
$.

(2)Rule for AA to process reply from WS: when an AA in a state $Invoking$ receives an \emph{invokeReply} reply message from its binding WS, it will be in a new state $Returned$ and send a \emph{notify} notification message to its WSOI.

$
\langle AA:Invoking\rangle, AA\lhd invokeReply(ops)@WS \nonumber\\
\xrightarrow[effect_{returnedAA}]{} \langle AA:Returned\rangle, \\ WSOI\lhd notify(AA=[State=Returned])@AA\nonumber
$.

Where $effect_{returnedAA}=setA(returnedAAU)$, and

$
returnedAAU=AA\{\\
\getAAQoS{AA}=qos_{AA}, \\ \getInputParametersFromAA{AA}=ips_{AA},\\
\getOutputParametersFromAA{AA}\\=ResultsFromWS(AA),\\ \getAAState{AA}=Returned,\\ \getWSOIId{AA}=\alpha_{cl},\\
\getWSFromAA{AA}=ws, \\ \getAAName{AA}=aaName\}
$,

where $ResultsFromWS(AA)$ denotes the results after a WS invocation.

The following is the transition rule for WS to accept an invocation from an AA: when a WS receives an \emph{invoke} request message from its binding AA, it will do some inner computations and send an \emph{invokeReply} reply message to its binding AA.

$
\langle WS:WSState\rangle, WS\lhd invoke(ips)@AA \nonumber\\
\rightarrow \langle WS:WSState'\rangle, AA\lhd invokeReply(ops)@WS\nonumber
$.

\textbf{Definition 4.17. (QoS-Aware WSO Behavior)}A system $S$ has QoS-Aware WSO Behavior, with respect to architecture as Fig.\ref{Fig.1} shows and actors WSOIM, WSOI, AAs, WSes, SS, if (1)for $C$ in $S$, states of actors in $C$ are just in accordance with definitions of states of actors; (2)for $C$ in $S$, messages in $C$ are just as Table.\ref{Table.1} define; (3)for every computation $\pi$ in $S$, $\pi$ obeys the transition rules defined above.

\textbf{Theorem 4.2. (System2Behavior)}If a system $S$ has QoS-Aware WSO Behavior as Definition 4.17 defines, Then $S$ is a QoS-Aware WSO System as Definition 4.16 defines.

\begin{proof}
By the states definition of actors WSOIM, WSOI, AAs, WSes, SS, message definitions as Table.\ref{Table.1} defined and the transition rule definitions above, we can see that(1)$\phi_{WSOI}(S)$ and $\phi_{WW}(S)$ hold; (2)for $C\in S$, if there is an undelivered request $WSOReq$, a new WSO instance $WSOI$ is assured to be created.
\end{proof}

\textbf{Theorem 4.3. (Service2Behavior)}If a system $S$ has QoS-Aware WSO Behavior as Definition 4.17 defines, then $S$ provides QoS-Aware WSO Service as defined in Definition 4.3.

\begin{proof}
By Theorem 4.1 and Theorem 4.2.
\end{proof}

\subsection{Behavior of QoS-Aware BookStore WSO}

The behavior of QoS-Aware BookStore WSO is embodied by the following transition rules.

(1)

$
\langle WSOIM:WSOIMState\rangle, \\ WSOIM\lhd wsoReq(\alpha_{cl}, WSO, qos, ips)@CA\nonumber\\
\xrightarrow[effect_{newWSOI}]{} \langle WSOIM:WSOIMState'\rangle,\\ SS\lhd select(qos, WSO)@WSOI\nonumber
$. $WSOI$ is a BookStore WSOI, and

(2)

$
\langle SS:SSState\rangle, \\SS\lhd select(qos, WSO)@WSOI \nonumber\\
\rightarrow \langle SS:SSState'\rangle, \\(WSOI\lhd selectReply\\\quad(WSOI=[State=Denied])@SS)
$. $WSOI$ is the BookStore WSOI, and

(3)

$
\langle WSOI:Waiting\rangle, \\ WSOI\lhd selectReply(WSOI=[State=Denied])@SS\\
\rightarrow \langle WSOI:Denied\rangle,\\ CA\lhd denied(WSO, qos)@WSOI
$. $WSOI$ is the BookStore WSOI.

or,

(1)

$
\langle WSOIM:WSOIMState\rangle, \\ WSOIM\lhd wsoReq(\alpha_{cl}, WSO, qos, ips)@CA\nonumber\\
\xrightarrow[effect_{newWSOI}]{} \langle WSOIM:WSOIMState'\rangle,\\ SS\lhd select(qos, WSO)@WSOI\nonumber
$. $WSOI$ is the BookStore WSOI, and

(2)

$
\langle SS:SSState\rangle, \\SS\lhd select(qos, WSO)@WSOI \nonumber\\
\rightarrow \langle SS:SSState'\rangle, \\WSOI\lhd selectReply\\\quad(WSOI=[State=Granted])@SS)
$. $WSOI$ is the BookStore WSOI, and

(3)

$
\langle WSOI:Waiting\rangle, WSOI \\ \lhd selectReply(WSOI=[State=Granted])@SS\\
\xrightarrow[effect_{grantedWSOI}]{} \langle WSOI:Granted\rangle, \\ CA\lhd granted(WSO, qos)@WSOI,\\
(\forall AA\in \getAAs{WSOI}) AA\lhd invoke()@WSOI
$. $WSOI$ is the BookStore WSOI, $AA \in \{Send List of Books, Receive Selected Books, \\Calculate the Price, Send Price of Books, \\Get Pays, Ship by Train or Ship by Air\}$, and

(4)

$
\langle AA:Preparing\rangle, \\ AA\lhd invoke()@WSOI \nonumber\\
\rightarrow \langle AA:Invoking\rangle, \\ WSOI\lhd invokeAck()@AA,  WS\lhd invoke(ips)@AA
$. $WSOI$ is the BookStore WSOI, $AA \in \{Send List of Books, Receive Selected Books, \\Calculate the Price, Send Price of Books, \\Get Pays, Ship by Train or Ship by Air\}$, $WS = getWS(AA)$, and

(5)

$
\langle WS:WSState\rangle, WS\lhd invoke(ips)@AA \nonumber\\
\rightarrow \langle WS:WSState'\rangle, AA\lhd invokeReply(ops)@WS\nonumber
$. $AA \in \{Send List of Books, Receive Selected Books, \\Calculate the Price, Send Price of Books, \\Get Pays, Ship by Train or Ship by Air\}$, $WS = getWS(AA)$, and

(6)

$
\langle AA:Invoking\rangle, AA\lhd invokeReply(ops)@WS \nonumber\\
\xrightarrow[effect_{returnedAA}]{} \langle AA:Returned\rangle, \\ WSOI\lhd notify(AA=[State=Returned])@AA\nonumber
$. $WSOI$ is the BookStore WSOI,

$AA \in \{Send List of Books, Receive Selected Books, \\Calculate the Price, Send Price of Books, \\Get Pays, Ship by Train or Ship by Air\}$, $WS = getWS(AA)$, and

(7)

$
\langle WSOI:\{Granted, Servicing\}\rangle, \\
WSOI\lhd invokeAck()@AA \\
\rightarrow \langle WSOI:Servicing\rangle
$. $WSOI$ is the BookStore WSOI, $AA \in \{Send List of Books, Receive Selected Books, \\Calculate the Price, Send Price of Books, \\Get Pays, Ship by Train or Ship by Air\}$, and

(8)

$
\langle WSOI:Servicing\rangle, \\
WSOI\lhd notify(AA=[State=Returned])@AA\nonumber\\
\xrightarrow[effect_{completedWSOI}]{} \langle WSOI:Completed\rangle, \\
CA\lhd completed(WSO, qos, ops)@WSOI
$. $WSOI$ is the BookStore WSOI, $AA \in \{Send List of Books, Receive Selected Books, \\Calculate the Price, Send Price of Books, \\Get Pays, Ship by Train or Ship by Air\}$.

Naturally, QoS-Aware BookStore WSO has QoS-Aware WSO behavior, and further provides QoS-Aware WSO Service according to Theorem 4.3.

\section{Conclusions and Future Works}

In this paper, we have discussed issues on QoS-aware WSO and design a typical QoS-aware WSO engine called QoS-WSOE. Mainly, a formal model of QoS-WSOE based on actor systems theory is established. In the formal model, a three-layered pyramidal structure is adopted to capture the requirements of the customers with a concept named QoS-Aware WSO Service, characteristics of QoS-WSOE with a concept named QoS-Aware WSO System, and behaviors of QoS-WSOE with a concept named QoS-Aware WSO Behavior and a fine relationship among these three layers is established. I hope this paper can be a guidance for implementing a real QoS-aware WSO Engine with correctness assurance.

In future, a more practical WSO engine including more properties, such as security, will be pursued.


\label{lastpage}

\end{document}